\documentclass{ws-ijmpa}

\usepackage[super,compress]{cite}

\begin{document}

\markboth{Massimiliano Rinaldi}{Acoustic entropy: massive case}

\catchline{}{}{}{}{}

\title{The entropy of an acoustic black hole in Bose-Einstein condensates: transverse modes as a cure for divergences. }

\author{Massimiliano Rinaldi} 

\address{Namur Center for Complex Systems (naXys) \\
University of Namur\\
8 Rempart de la Vierge, B-5000, Belgium
\footnote{mrinaldi@fundp.ac.be}}

\maketitle

\begin{abstract}

We consider the entropy associated to the phonons generated via the Hawking mechanism in a sonic hole in a Bose-Einsten condensate. In a previous paper, we looked at the (1+1)-dimensional case both in the hydrodynamic limit and in the case when high-frequency dispersion is taken in account. Here, we extend the analysis, based on the 't Hooft brick wall model, by including transverse excitations. We show that they can cure the infrared divergence that appears in the (1+1)-dimensional case, by acting as an effective mass for the phonons. In the hydrodynamic limit, where high-frequency dispersion is neglected, the ultraviolet divergence remains. On the contrary, in the dispersive case the entropy not only is finite, but it  is completely fixed by the geometric parameters of the system.
\end{abstract}

\ccode{04.70.Dy, 04.62.+v, 03.75.Gg}

\newcommand{\be}{\begin{equation}}
\newcommand{\ee}{\end{equation}}
\newcommand{\bea}{\begin{eqnarray}}
\newcommand{\eea}{\end{eqnarray}}
\newcommand{\non}{\nonumber}


\section{Introduction}


\noindent Direct observations of the Hawking radiation \cite{Hawking} are extremely difficult, mainly because   $\hbar$ and $c^{-1}$ are very small and this leads to a temperature of several orders of magnitude lower than the cosmic microwave background for a black hole of a solar mass. However, in condensed matter physics there are systems able to reproduce curved spacetime configurations, such that the speed of light is effectively replaced by the speed of sound waves. These are described by an equation of motion formally identical to the relativistic Klein-Gordon equation on a curved background \cite{unruh}. In this context, an irrotational fluid accelerated from subsonic to supersonic speed generates a thermal flux of  phonons with a spectrum analogous to the one associated to the Hawking radiation emitted by a gravitational black hole \cite{cimento,reviewbec}. These systems, known as acoustic black holes (or dumb/sonic holes), have attracted a great attention in the past years, especially because they might be soon realized in a laboratory.

The analogy between astrophysical and acoustic black holes has been strengthened thanks to the careful analysis of the correlation functions of the modes generated via the Hawking mechanism   near the acoustic horizon, both analytically \cite{cimento,reviewbec,balb-carus-fabbri,mayo-fabbri,mayo-fabbri-rinaldi,analyticBEC} and numerically \cite{numericBEC}. The non-trivial structure of these correlation functions leads  to ask what are the features of the entanglement entropy associated to these modes. For astrophysical black holes, it is well known that the area of the horizon is proportional to the thermodynamical entropy \cite{bek}. However, for acoustic black holes such a formula does not seem to exists. Nevertheless, it is possible to calculate the entanglement entropy between the pairs of phonons created via the Hawking mechanism. In Ref.\ \citen{acentropy}, we looked at this problem by considering the simplest system, namely a (1+1)-dimensional acoustic black hole in a  Bose-Einstein Condensate (BEC). In this work, we used the so-called brick wall method, elaborated by 't Hooft several years ago and applied to gravitational black holes  \cite{thooft,Muko,reviewentropy}.  At first sight, such a system seems trivial. In the realm of gravity, it is known that (1+1)-dimensional black holes are conformally invariant, and their entanglement entropy must be independent of the only characteristic scale of the black hole, namely its mass. In fact, in these systems the entropy is simply given by $S\sim \ln (L/\epsilon)$, where $L$ and $\epsilon$ are the infrared (IR) and the ultraviolet (UV) cutoff respectively \cite{entangl,callan}. For black holes in four dimensions, conformal invariance is broken, and the entanglement entropy turns out to scale as the area of the horizon, just like the thermal entropy. In the acoustic (1+1)-dimensional acoustic black holes studied in Ref.\ \citen{acentropy}, we broke conformal invariance by introducing high frequency dispersion and we found non-trivial corrections to the entropy.

Technically speaking, there is no rigorous proof  that the entanglement entropy is the same as the brick wall entropy. However, there are strong evidences that this is the case, at least in the condensed matter systems considered here. To begin with, in Ref.\  \citen{acentropy} we showed that the two kinds of entropy are exactly the same in the conformal case, i.e. when no dispersion is taken in account. The introduction of high frequency dispersion leads to a correction to the leading term that is the same as the one found in spin-1/2 Heisenberg XX chains, using the ``replica trick'', see e.g. Ref.\ \citen{calabrese}. Finally, the results in Ref.\ \citen{acentropy} are in agreement with the ones found in Ref.\ \citen{giovanazzi}, where the entanglement entropy was calculated using the density matrix $\rho$ and the usual formula $S=-{\rm Tr} \rho \ln \rho$ in a (1+1)-dimensional degenerate ideal Fermi gas with a sonic event horizon. Concerning astrophysical black holes, it has been shown that the two entropies are both proportional to the horizon area but they show a UV divergence that can be cured with the introduction of a cutoff around the Planck scale, see e.g. Ref.\ \citen{brustein}, or a modification of the dispersion relation \cite{UVentropy,remo}. Therefore it is possible that a proper regularization scheme makes entanglement entropy the same as the brick wall one. Such a suggestion is explored in Ref.\ \citen{Demers}, where the brick wall entropy is calculated for a Reissner-Nordstr\"om black hole and the divergences are regularized with the Pauli-Villar method. The authors found logarithmic corrections that are the same as the one-loop corrections to the Bekenstein entropy calculated with other methods. Later, is was realized that these corrections are the same as the one found with the ``replica trick'' whenever the background has vanishing Ricci scalar \cite{reviewentropy}.

In  (1+1)-dimensional sonic holes in BEC, the UV cutoff is fixed by a specific dispersion function at high frequency that  introduces a differential operator of order four in the mode equation for the phonon field  outside the horizon, in analogy with certain gravitational models endowed with modified dispersion relations \cite{mdr}. In addition, when transverse phonon excitations are taken in account, they act as an effective mass that can remove IR divergences. Therefore, we expect that the entropy is completely determined by the microscopic parameters of these sonic holes, in opposition to the conformal case, where it depends only on the linear size and on an arbitrary cut-off.

In the present work, we aim to prove this by extending the results of Ref.\ \citen{acentropy} to the case when  transverse modes are excited. Typical laboratory BEC systems are made by a flow of ultracold atoms (e.g. Rubidium) modulated along a waveguide. If no transverse excitations are allowed, the system is essentially (1+1)-dimensional. However, transverse modes can be taken in account experimentally  \cite{stringari}.

The plan of the paper is the following. In Sec.\ 2 we lay down the basic equations that describe the acoustic black hole with the inclusion of transverse modes. In Sec.\ 3 we calculate the entropy in the hydrodynamic limit in the presence of transverse excitations. In Sec.\ 4 we compute the entropy by including high frequency dispersion and we discuss our results in Sec.\ 5.

\section{The sonic black hole}

In the dilute gas approximation \cite{stringari}, the BEC can be described by an operator $\hat \Psi$ that obeys the equation
\bea
i\hbar \partial_{t}\hat\Psi=\left( -{\hbar^{2}\over 2m} \vec{\nabla}^{2}+V_{\rm ext} + g\hat\Psi^{\dagger}\hat\Psi\right)\hat\Psi\ ,
\eea
where $m$ is the mass of the atoms, $g$ is the non-linear atom-atom interaction constant, and $V_{\rm ext}$ is the external trapping potential. The wave operator satisfies the canonical commutation relations $[\hat\Psi(t,\vec{x}),\hat\Psi(t,\vec{x}')]=\delta^{3}(\vec{x}-\vec{x}')$. To study linear fluctuations, one substitutes $\hat\Psi$ with $\Psi_{0}(1+\hat\phi)$ so that $\Psi_{0}$ solves the Gross-Pitaevski equation 
\bea\label{GP}
i\hbar \partial_{t}\Psi_{0}=\left( -{\hbar^{2}\over 2m} \vec{\nabla}^{2}+V_{\rm ext} + gn\right)\Psi_{0}\ ,
\eea
and the fluctuation $\hat\phi$ is governed by the Bogolubov-de Gennes equation
\bea\label{bdg}
i\hbar \partial_{t}\hat\phi=-{\hbar^{2}\over 2m}\left(\vec{\nabla}^{2}+2{\vec{\nabla}\Psi_{0}\over \Psi_{0}}\vec{\nabla}\right)\hat\phi+mc^{2}(\hat\phi+\hat\phi^{\dagger}),
\eea
where $c=\sqrt{gn/m}$ is the modulus of the speed of sound, and $n=|\Psi_{0}|^{2}$ is the number density. The commutation relations for $\hat \phi$ are
\bea\label{normcon}
[\hat\phi(t,\vec{x}),\hat\phi^{\dagger}(t,\vec{x}')]=n^{-1}\delta^{(3)}(\vec x-\vec x').
\eea

We now focus on a one-dimensional flow  of BEC that, conventionally, moves along the $x$-direction from right to left with constant velocity $\vec v=(-v,0,0)$ and fixed $n$, while $\vec c=(-c(x),0,0)$ smoothly varies in such a way that at the surface $x=0$ we have $c(x) = v$. Conventionally, we choose $c(x)< v(>v)$ for $x<0 (>0)$, thus we have a sonic horizon at $x=0$, with a supersonic region on the left and a subsonic one on the right. With this choice, $v$ and $c(x)$ are positive numbers.

It is possible to adjust $c(x)$ provided one allows for a varying $g$, such that the quantity $gn+V_{\rm ext}$ remains constant in space \cite{numericBEC}. In this way, Eq.\ (\ref{GP}) admits the plane-wave solution $\Psi_{0}=\sqrt{n}\exp (i \vec k_{0}\cdot \vec x-i\omega_{0}t)$, where  $\vec k_{0}$ is related to the condensate velocity through the relation  $\vec v=\hbar \vec k_{0}/m$. To study the dynamics of $\hat \phi$, it is convenient to consider  the so-called density-phase representation defined by 
 \bea
 \hat\phi={\hat n\over 2n}+{i\hat\theta\over \hbar}\ ,
 \eea
 along the lines of Ref.\ \citen{balb-carus-fabbri}. By choosing a one-dimensional flowing condensate, we can write the (3+1)-dimensional acoustic metric  in the form \cite{cimento}
 \bea\label{acmetric}
ds^{2}&\!=\!&{n\over mc(x)}\left[-(c^{2}(x)-v^{2})dt^{2}-2vdxdt+dx^{2}+dy^{2}+dz^{2}\right],
\eea
which represents an acoustic black hole in $(1+1)$-dimensions, implemented by two asymptotically flat directions. In fact, this metric has an event horizon located where $c(x)=v$ and its structure is the same of the Painlev\'e-Gullstrand line element \cite{PG}, up to the conformal factor $n/(mc)$. Normally, one discards transverse directions by considering only $x$-dependent phonon fields $\Psi$ and by neglecting the $(y,z)$ metric components. In this way, one is left with a genuine $(1+1)$-dimensional field propagating on a $(1+1)$-dimensional metric. However, this is only an approximation of the real system. In fact, the actual physical model is an elongated cloud of ultracold atoms trapped by means of magnetic fields and moving with constant speed along the $x$-axis, where transverse mode can be excited \cite{numericBEC}. Here we propose to take in account these additional transverse degrees of freedom by assuming that the  phonon components are $(3+1)$-dimensional fields of the form
 \bea
 \hat\theta&=&\hat\theta (x,t)\,\exp(ik_{y}y+ik_{z}z)\ ,\\
  \hat n&=&\hat n (x,t)\,\exp(ik_{y}y+ik_{z}z)\ .
 \eea
 If we now expand the operators $\hat \theta$ and $\hat n$ into sums of creation and annihilation operators, i.\ e.\ in the form
 \bea
 \hat n=\sum_{j} (\hat a_{j}\, \tilde n_{j} + {\rm h.c.}),\quad \hat \theta=\sum_{j} (\hat a_{j}\, \tilde \theta_{j} + {\rm h.c.})\ ,
 \eea
 we then find, from Eq.\ \eqref{bdg} implemented by the plane wave-expansion of Eq.\ \eqref{GP}, the two coupled equations
 \bea\non
 0&=&(\partial_{t}+v\partial_{x})\tilde\theta-{\hbar^{2}\over4mn }\partial_{x}^{2}\tilde n+\left({\hbar^{2}\over 4mn\lambda^{2}}+{\hbar c\over n\xi}\right)\tilde n\ ,\\
  0&=&(\partial_{t}+v\partial_{x})\tilde n+{n\over m}\partial_{x}^{2}\tilde\theta-{n\over m \lambda^{2}}\tilde \theta\ ,\label{dyneqs}
 \eea
 where  $\lambda^{-2}=k_{y}^{2}+k_{z}^{2}$. The function $\xi=\hbar/(mc)$ is known as the healing length, which sets roughly the scale at which the model breaks down \cite{stringari}. 
The commutation relations for  $\hat \theta$ and $\hat n$ are $[\hat n(t,x),\hat\theta(t,x')]=i\hbar\delta(x-x')$, from which one obtains the normalization of the modes, fixed by the relation \cite{mayo-fabbri}
\bea\label{norm}
\int dx\left( \tilde n_{j} \tilde\theta^{*}_{j'}- \tilde n_{j}^{*} \tilde\theta_{j'}  \right)=i\delta_{ij}.
\eea
 
In the expressions \eqref{dyneqs}, we see that the effects of the  transverse modes appear as an effective mass, that can be quantized in the case when the $(y,z)$ range is bounded or periodic. Formally, Eqs.\ \eqref{dyneqs} can be interpreted also as the dynamical equations for a scalar field of mass $\lambda^{-2}$ propagating only on the $(t,x)$ sector of the acoustic metric \cite{parelast}.   In fact, one can show that these equations can be found also by expanding the massive 2-dimensional Klein-Gordon operator (together by the appropriated high-frequency dispersion operator) on the $(t,x)$ sector of the metric \eqref{acmetric}.

We wish to stress, however, that, although the system is mathematically described by  (1+1)-dimensional massive fields, the physics is in fact 4-dimensional. The dynamical equations \eqref{dyneqs} depend on $(x,t)$ only because of the particular geometry chosen for the background, but the ``mass'' term $\lambda^{-2}$ keeps track of the transverse directions. Indeed, it is this term that leads to expressions for the entropy depending on the cross area of the beam, as we will shortly see.

In what follows, we will consider  phononic modes with wavelength much smaller than $\xi$.  Note that, in our settings, $\xi$ is not constant, as it depends on the (varying) speed of sound. The limit $\xi=0$ corresponds to the so-called hydrodynamic limit. Within this regime, one can show that the two equations above can be combined together into
 \bea\label{eom}
\partial_{x}^{2}\tilde\theta(x)&=&\left[(\partial_{t}+v\partial_{x}){1\over c(x)^{2}}(\partial_{t}+v\partial_{x})+{1\over\lambda^{2}}\right]\tilde\theta(x),\\
\tilde n(x)&=&-{4nm\lambda^{2}\over \hbar^{2}+4m^{2}c^{2}\lambda^{2}}(\partial_{t}+v\partial_{x})\tilde\theta(x)\ .\label{ntheta}
 \eea
 The normalization condition for $\tilde \theta$ follows from equations \eqref{norm} and \eqref{ntheta} and reads
 \bea
 i\delta_{ij}=\int dx{4nm\lambda^{2}\over \hbar^{2}+4m^{2}c^{2}\lambda^{2}}\left[ \tilde\theta_{j'}(\partial_{t}+v\partial_{x})\tilde \theta_{j}^{*} - \tilde\theta_{j'}^{*}(\partial_{t}+v\partial_{x})\tilde \theta_{j}  \right].
 \eea
 In the massless limit, $\lambda\rightarrow \infty$, we obtain the same normalization as in ref. [\citen{mayo-fabbri}].

\section{Brick wall and entanglement entropy in the hydrodynamic limit}

In this section, we study the equation (\ref{eom}) and use the 't Hooft brick-wall method to compute the entropy \cite{thooft}. To begin with, we assume that the field $\tilde\theta$ is stationary, and we set $\tilde\theta=\theta(x,\lambda)\exp (i\omega t)$. Then, Eq.\ (\ref{eom}) can be written in the form
\bea\label{eomf}
\theta''(x,\lambda)+2A(x)\theta'(x,\lambda)+B(x,\lambda)\theta(x,\lambda)=0\ ,
\eea
where
\bea
A(x)={i\omega v+v^{2}\frac{c'}{c}\over c^{2}-v^{2}},\quad B(x,\lambda)={\omega^{2}-2i\omega v{c'\over c}-{c^{2}\over\lambda^{2}}\over c^{2}-v^{2} }\ .
\eea
For notational simplicity, from now on the $x$-dependence of $c(x)$ will be understood. To apply the WKB formula, we make the substitution
\bea\label{wkbsub}
\theta(x,\lambda)={N\over\sqrt{g(x,\lambda)}}\exp\left[{i\over\hbar}\int f(x,\lambda)dx\right]\ ,
\eea
where
\bea
f(x,\lambda)=g(x,\lambda)+i\hbar A(x),
\eea
and $N$ is a normalization constant.
In this way, Eq.\ (\ref{eomf}) assumes the WKB form
\bea\label{eomwkb}
{3\over 4}{g'^{2}\over g^{2}}-{g^{2}\over \hbar^{2}}-{1\over 2}{g''\over g}+P(x)+{V(x,\lambda)\over \hbar^{2}}=0\ .
\eea
where
\bea\non
P(x)&=&{v^{2}(3c^{2}-2v^{2})\over (c^{2}-v^{2})^{2}}\left(c'\over c\right)^{2}-{v^{2}\over (c^{2}-v^{2})}{c''\over c}\ ,\\
V(x,\lambda)&=&{c^{2}\hbar^{2}\over \lambda^{2}}\left[{\omega^{2}\lambda^{2}\over (c^{2}-v^{2})^{2}}-{1\over (c^{2}-v^{2})}\right]\ .\label{V}
\eea
The mode density number is obtained by integrating $f(x,\lambda)$ over the near-horizon region. If we assume that the transverse momenta $k_{y}$ and $k_{z}$ are quantized, we also need to sum over all their possible values, in analogy with the sum over the angular momenta in the gravitational $(3+1)$-dimensional case. 

The WKB analysis proceeds by substituting $g=g_{0}+\hbar^{2}g_{2}+\hbar^{4}g_{4}+\ldots$ in the above equation and by collecting the terms with the same power of $\hbar$. The lowest  order is  
\bea\label{lowestwkb}
g_{0}(x,\lambda)=\pm\sqrt{V(x,\lambda)}\ ,
\eea
so the lowest WKB order solution to Eq.\ (\ref{eomf}) reads
\bea
f_{0}(x,\lambda)=\pm\sqrt{V(x,\lambda)}+i\hbar A(x)\ .
\eea

We find convenient to sum first over the transverse momenta. By turning the infinite sum into an integral we have

\bea\label{effe0}
f_{0}(x)&=& {c\,\hbar\,\omega\over c^{2}-v^{2}} {\sigma\over (2\pi)^{2}}\int dk_{y}dk_{z}\sqrt{1-Z^{2}(k_{y}^{2}+k_{z}^{2})}+i\hbar A(x),
\eea
where we set
\bea
Z^{2}={c^{2}-v^{2}\over \omega^{2}},
\eea
and $\sigma/(2\pi)^{2}$ is the surface spanned by the transverse oscillations. The sign has been fixed so that we recover the known case for $k_{y},k_{z}=0$. By integrating in the region where the square root is real, we find
\bea
f_{0}(x)={\hbar\omega\over 2\kappa x}+{\sigma \hbar \omega^{3}\over 24\pi v \kappa^{2}x^{2}}.
\eea
To obtain this formula, we used the approximation $c=v+\kappa x$, where $\kappa=(dc/dx)_{x=0}$ is the analog of the surface gravity of a black hole. This is justified by the hypothesis that we want to integrate the mode number in the near-horizon region, where $c(x)$ is linearized. The same approximation is usually applied to the $(3+1)$-dimensional gravitational black hole. 

We now compute the entanglement entropy by integrating over the number of modes with energy $E=\hbar\omega$ living in a segment with one endpoint placed at an arbitrarily small distance $x=\epsilon$ from the horizon and the other at $x=L$, where $L$ can be identified with the length of the region where the near-horizon approximation is valid. It is easy to show that $L=v/\kappa$ is a good estimate for the length of this region, provided $\kappa$ is a smooth function of $x$. We find
\bea\label{xmodes}
n(E)={E\over 2\pi \hbar \kappa}\ln \left(L\over \epsilon\right)+{\sigma E^{3}\over 24 \pi^{2}L\epsilon \hbar^{3}\kappa^{3}}.
\eea

We now recall that the free energy and the entropy associated to massless spin-0 particles are given respectively by
\bea\label{freeen}
F&=&-\int_{0}^{\infty}{n(E)dE\over (e^{\beta E}-1)}, \\
S&=&\beta^{2}{dF\over d\beta}\ ,\label{entro} \label{entropy}
\eea
where $\beta=(k_{\rm B}T)^{-1}=2\pi/(\hbar\kappa)$ is the inverse of the (black hole) temperature. With these formulae, we find that the entropy reads
\bea\label{Shydro}
S_{\lambda}={1\over 12}\ln \left(L\over \epsilon\right)+{\sigma\over 720 \pi L\epsilon}
\eea
We see that the transverse modes contributes to the entropy with a geometric term that still diverges in the UV ($\epsilon\rightarrow 0$). If we allow for a transverse section with a radius of the order of $L$, $\sigma=\pi  L^{2}$, we see that this contribution is negligible for $L\gg \epsilon$.
Had we not taken in account these, we would have found \cite{acentropy}
\bea
S_{\lambda\rightarrow\infty}={1\over 6}\ln \left(L\over \epsilon\right).
\eea
One might be worried because the expression \eqref{Shydro} does not converge to this result for $\sigma\rightarrow 0$. This is not so surprising as the limit $\lambda\rightarrow\infty$ and the integral in Eq.\ \eqref{effe0}
 do not commute. The same situation occurs in the usual 't Hooft brick wall model: if one does not integrate over the angular momentum by choosing, for example, $\ell=0$ the result is radically different and one does not even recover the proportionality between horizon area and entropy. The conclusion of this section is that the transverse mode contributes to the entropy but are not able to cure the ultraviolet divergence in the hydrodynamic limit. The situation is quite different when dispersion is taken in account, as we show in the next section.

\section{Dispersive case}
We now consider the dispersive case, and evaluate the contribution to the entropy given by  the transverse excitations. In order to do so, it is more convenient to write the Eq. (\ref{bdg}) by using the expansion $\hat\phi=\sum_{j}[\hat a \tilde\phi + \hat a^{\dagger}\tilde\varphi^{*}]$. By assuming, as above, that $\tilde\phi=\phi(x)\exp (i\omega t)\exp (ik_{y}y+ik_{z}z)$ and that $\tilde\varphi=\varphi(x)\exp (i\omega t)\exp (ik_{y}y+ik_{z}z)$, we obtain the two coupled equations
\bea\non
(-\omega+iv\partial_{x})\phi+{c\xi\over 2}\partial_{x}^{2}\phi-{c\xi\over 2\lambda^{2}}\phi-{c\over \xi}(\phi+\varphi)=0,\\
(\omega-iv\partial_{x})\varphi+{c\xi\over 2}\partial_{x}^{2}\varphi-{c\xi\over 2\lambda^{2}}\varphi-{c\over \xi}(\phi+\varphi)=0 .\label{dispcase}
\eea
The normalization conditions following from eq.\ \eqref{normcon} are\cite{mayo-fabbri-rinaldi}
\bea
n\hbar \int dx \left( \phi_{j}\phi_{j'}^{*}-\varphi_{j}^{*}\varphi_{j'}  \right)=\delta_{jj'}.
\eea
By following Ref.\ \citen{acentropy}, we decouple these equations in momentum space. By defining 
\bea
\psi^{\pm}(x)=\phi(x)\pm\varphi(x), \,\,\,\, \tilde\psi^{\pm}(p)=\int{dx e^{-ipx}\over\sqrt{2\pi}}\psi^{\pm}(x),
\eea
we find
\bea\label{modeP}
\hat c^{2} \tilde \Psi^{+}(p)&=&V(p,\lambda)\tilde  \Psi^{+}(p)\ ,\\\non\\
\tilde \Psi^{-}(p)&=&{2m\lambda^{2}(\omega-pv)\over \hbar (p^{2}\lambda^{2}+1)}\tilde \Psi^{+}(p),
\eea
where
\bea
V(p,\lambda)= {\lambda^{2}(\omega-pv)^{2}\over \lambda^{2}p^{2}+1}-{\hbar^{2}\over 4m^{2}}\left(p^{2}\lambda^{2}+1\over \lambda^{2}\right).
\eea
Eq.\ (\ref{modeP}) correctly reduces to the one found in Ref.\ \citen{acentropy} when $\lambda\rightarrow\infty$ and we can solve it in the near-horizon approximation, i.e. when $\hat c=v+i\kappa \partial_{p}$. In analogy to what we did in the hydrodynamic limit, we use a WKB procedure and replace the expression
\bea
\psi^{+}(p)={N\over{\sqrt{f(p)}}}\exp\left[{i\over \hbar}\int^{p'} dp \left(f(p)+{v\hbar\over\kappa}\right)\right].
\eea
in Eq.\ (\ref{modeP}) to find
\bea
{3\over 4}\left(f'\over f\right)^{2}-{f''\over 2f}-{f^{2}\over \hbar^{2}}+{V\over \kappa^{2}}=0,
\eea
where the prime here denotes differentiation with respect to $p$. Note that the term $v\hbar/\kappa$ is not present in the momentum representation used in Ref.\ \citen{acentropy}. This is due to the fact that here we use the $(\phi,\, \varphi)$ representation while in the hydrodynamic limit of Ref.\ \citen{acentropy} we use the density-phase representation $(n_{1},\, \theta_{1})$.

By keeping the lowest order only, we find the solution
\bea\label{dispwkb}
f(p,\lambda)={v\hbar\over \kappa}\left[1\pm\sqrt{{\lambda^{2}p^{2}\over \lambda^{2}p^{2}+1}\left(1-{\omega \over vp}\right)^{2}-{\xi_{0}^{2}\over 4}\left(\lambda^{2}p^{2}+1\over \lambda^{2}\right)}\,\,\right]\ ,
\eea
where $\xi_{0}=\xi(x=0)$ is the healing length at the horizon. By taking the hydrodynamic limit $\xi_{0}\rightarrow 0$, and by computing the number of modes, we see that, in order to recover Eq.\ (\ref{xmodes}) we must consider only the sign $+$, a choice that will be understood in the following. For $\xi_{0}\neq 0$, we can check that, in the limit $\lambda\rightarrow \infty$, this equation becomes the same as Eq.\ (21) of Ref.\ \citen{acentropy} as expected. However, the role of $\lambda$ turns out to be crucial. In fact, we note that Eq.\ (\ref{dispwkb}) is no longer singular in the limit $p\rightarrow 0$. This means that in the computation of the number of modes we do not have to worry about the low-momentum modes, which had to be eliminated by a IR cut-off in the calculations of Ref.\ \citen{acentropy}. Therefore, we see that the effect of the transverse modes is to naturally regularize the infrared divergence.

The number of modes with a given energy $E=\hbar\omega$ follows from the formula 
\bea\label{modeno}
n(E)={1\over \pi\hbar}\int_{p_{\rm min}}^{p_{\rm max}}\sum_{\lambda}f(p,\lambda)dp, 
\eea
where the interval of integration is determined by the condition that the square root in $f(p,\lambda)$ is real.
Before integrating over $p$, we need to calculate the sum over the possible values of $\lambda$. As we did in the hydrodynamic limit, we turn the sum over $\lambda$ into an integral so that
\bea
n(E)={v\over \pi \kappa}\int_{0}^{\bar p}dp\left[1+{\sigma \xi_{0}p^{3}\over 8\pi}\left({4\over 3}H^{3/ 2}E_{F}\left({\sqrt{H-1}\over \sqrt{2H}},\sqrt{2}\right)-{2\over 3}\sqrt{H^{2}-1}\right)\right],
\eea
where
\bea
H={2\over \xi_{0}p^{2}}\left({\omega\over v}-p\right),
\eea
and $E_{F}$ is the incomplete elliptic integral of first kind. The range of integration $[0,\bar p\,]$ coincide with $H\in [1,\infty[$. Exact integration is not possible, but, by noting that the integral is dominated by large values of $H$, approximations can be made and the mode density number can be written as
\bea
n(E)={E\over \pi\kappa\hbar}+{L\sigma\over \xi^{3}_{0}}\sum_{j}\alpha_{j}\left(E\xi_{0}\over \hbar v\right)^{j}
\eea
where, again, we set $L=v/\kappa$ as the size of the near-horizon region, and $\alpha_{j}$ are numerical coefficients. The index $j$ run over integers greater than 4 except for the first two terms of the sum, for which $j=5/2$ and $j=3$. In any case the terms of the sum are subdominant in the calculation of the entropy. In fact, by using the formulae \eqref{freeen} and \eqref{entropy}, we find that the entropy has the form
\bea
S={1\over 6}+{\sigma\over \xi_{0}^{2}}\left[\tilde\alpha_{5/2}\left(\xi_{0}\over L\right)^{5/2} + \tilde\alpha_{3}\left(\xi_{0}\over L\right)^{3} +\sum_{j=5}^{\infty}\tilde\alpha_{j}\left(\xi_{0}\over L\right)^{j}\right],
\eea
where the $\tilde \alpha_{j}$'s differ from  the $\alpha_{j}$'s by unimportant numerical factors. The important result instead is that the leading term is a constant and the logarithmic dependence in Eq.\  \eqref{Shydro} has disappeared. The correction terms are all small because of the condition that the healing length at the horizon $\xi_{0}$ is  much smaller than the size of the  near-horizon region $L$. Again, if we allow for a transverse section with radius of the order of the near horizon region, we see that all the term in the sum have the form $(\xi_{0}/L)^{k}$ with $k>0$.

\section{Discussion}

\noindent Let us briefly summarize the main results. In the hydrodynamic limit, we computed the entanglement entropy by taking in account transverse excitations. As expected, these act as an infrared cutoff, and the corrections to the usual logarithmic term $(1/6) \ln(L/\epsilon)$ have the form  $\sigma/ L\epsilon$. The presence of the transverse modes does not help in curing the UV divergence encoded in the limit $\epsilon\rightarrow 0$ and that represents the effect of modes boundlessly piling up near the horizon, as a result of the lack of any ultraviolet completion of the field theory.

In this perspective, we find very interesting the results of Sec.\ 4, where  dispersion effects are taken in account. For a BEC the dispersion relation is modified at high frequency according to the function 
\bea
(\omega-vp)^{2}=c^{2}\left(p^{2}+{\xi^{2}p^{4}\over 4}\right)\ .
\eea
In the case when transverse directions are taken in account, the above formula is generalized to
\bea
(\omega-vp)^{2}=c^{2}\left({1\over \lambda^{2}}+p^{2}\right)\left(1+{\xi^{2}p^{2}\over 4}+{\xi^{2}\over 4\lambda^{2}}\right)\ ,
\eea
as it can be easily found by replacing plane wave solutions to $\phi$ and $\varphi$ in Eqs.\ \eqref{dispcase}, see Ref.\ Ê\citen{mayo-fabbri-rinaldi}. In Ref.\ \citen{acentropy} we have found that the modified dispersion relation acts as an ultraviolet cutoff and the entanglement entropy receives a correction that depends on the healing length. The leading term is however still arbitrary, as it depends on the largest wavelength allowed in the system. This reflects the lack of a physical infrared cutoff that can be provided for by taking in account the effects of transverse modes. As we have seen in Sec.\ 4, the density of modes turns out to be finite in the near-horizon region, thus revealing a surprising IR/UV mixing: the infrared cutoff $\lambda$ prevents the divergence of the density of modes near the horizon. This has a dramatic consequence as the leading term of the entropy becomes a constant and the corrections are uniquely fixed by geometric parameters of the system: the healing length $\xi_{0}$, the extension of the near-horizon region $L$, and the cross section $\sigma$. One challenging issue is to verify that the entanglement entropy found here is the same as the one calculated by other methods. We hope to report soon on this topic.

\section*{Acknowledgements}

M.R. is a postdoctoral fellow of the Joint Research Action in Cosmology (ARC-COS) of the Louvain Academy, grant No. ARC 11/15-040. The author wishes to thank R.\ Balbinot and I.\ Carusotto for helpful discussions.

\end{document}